\documentclass[global,twocolumn,abstract]{svjour}

\usepackage{epsfig}

\begin{document}

\authorrunning{A.\,Nikolaev \textit{et al.}}
\titlerunning{
Determination of the $\eta$ mass with the Crystal Ball at MAMI-B}
\title{
Determination of the \boldmath $\eta$ \unboldmath mass with the Crystal Ball at MAMI-B}

\author{
The Crystal Ball at MAMI, TAPS and A2 Collaborations\\
\\
A.\,Nikolaev\inst{1}\and
P.\,Aguar-Bartolom\'{e}\inst{\,2}\and
J.\,Ahrens\inst{2}\and
J.\,R.\,M.\,Annand\inst{\,3}\and
H.\,J.\,Arends\inst{2}\and
R.\,Beck\inst{1,}\thanks{E-mail: beck@hiskp.uni-bonn.de}\and
V.\,Bekrenev\inst{4}\and
B.\,Boillat\inst{5}\and
A.\,Braghieri\inst{6}\and
D.\,Branford\inst{7}\and
W.\,J.\,Briscoe\inst{8}\and
J.\,W.\,Brudvik\inst{9}\and
S.\,Cherepnya\inst{10}\and
R.\,Codling\inst{3}\and
M.\,Dehn\inst{2}\and
E.\,J.\,Downie\inst{3}\and
L.\,V.\,Fil'kov\inst{10}\and
D.\,I.\,Glazier\inst{7}\and
R.\,Gregor\inst{11}\and
E.\,Heid\inst{2}\and
D.\,Hornidge\inst{12}\and
O.\,Jahn\inst{2}\and
A.\,Jankowiak\inst{2}\and
K.-H.\,Kaiser\inst{2}\and
V.\,L.\,Kashevarov\inst{10}\and
R.\,Kondratiev\inst{13}\and
M.\,Korolija\inst{14}\and
M.\,Kotulla\inst{11}\and
D.\,Krambrich\inst{2}\and
B.\,Krusche\inst{5}\and
M.\,Lang\inst{1}\and
V.\,Lisin\inst{13}\and
K.\,Livingston\inst{3}\and
U.\,Ludwig-Mertin\inst{2}\and
S.\,Lugert\inst{11}\and
I.\,J.\,D.\,MacGregor\inst{3}\and
D.\,M.\,Manley\inst{15}\and
M.\,Martinez-Fabregate\inst{2}\and
J.\,C.\,McGeorge\inst{3}\and
D.\,Mekterovic\inst{14}\and
V.\,Metag\inst{11}\and
B.\,M.\,K.\,Nefkens\inst{9}\and
R.\,Novotny\inst{11}\and
R.\,O.\,Owens\inst{3}\and
P.\,Pedroni\inst{6}\and
A.\,Polonski\inst{13}\and
S.\,N.\,Prakhov\inst{9}\and
J.\,W.\,Price\inst{9}\and
A.\,Reiter\inst{2,3}\and
G.\,Rosner\inst{3}\and
M.\,Rost\inst{2}\and
T.\,Rostomyan\inst{5,6}\and
S.\,Schumann\inst{1,2}\and
D.\,Sober\inst{16}\and
A.\,Starostin\inst{9}\and
I.\,Supek\inst{14}\and
C.\,M.\,Tarbert\inst{7}\and
A.\,Thomas\inst{2}\and
M.\,Unverzagt\inst{1,2}\and
Th.\,Walcher\inst{2}\and
D.\,P.\,Watts\inst{7}\and
F.\,Zehr\inst{5}}

\institute{
Helmholtz-Institut f\"ur Strahlen- und Kernphysik, University Bonn, Bonn, Germany \and
Institut f\"ur Kernphysik, University Mainz, Mainz, Germany \and
Department of Physics and Astronomy, University of Glasgow, Glasgow, United Kingdom \and
Petersburg Nuclear Physics Institute, Gatchina, Russia \and
Institut f\"ur Physik, University Basel, Basel, Switzerland \and
INFN Sezione di Pavia, Pavia, Italy \and
School of Physics, University of Edinburgh, Edinburgh, United Kingdom \and
Center for Nuclear Studies, The George Washington University, Washington, D.C., USA \and
University of California Los Angeles, Los Angeles, California, USA \and
Lebedev Physical Institute, Moscow, Russia \and
II. Physikalisches Institut, University Gie\ss en, Gie\ss en, Germany \and
Mount Allison University, Sackville, NB, Canada \and
Institute for Nuclear Research, Moscow, Russia \and
Rudjer Boskovic Institute, Zagreb, Croatia \and
Kent State University, Kent, Ohio, USA \and
The Catholic University of America, Washington, D.\,C., USA
}

\date{Received: date / Revised version: }       


\maketitle

\begin{abstract}
A new precise determination of the $\eta$ meson mass is presented. It is based on a measurement of the threshold for the $\gamma p \to p \eta$ reaction using the tagger focal-plane microscope detector at the MAMI-B facility in Mainz. The tagger microscope has a higher energy resolution than the standard tagging spectrometer and, hence, allowed an improvement in the accuracy compared to the previous $\eta$ mass measurement at MAMI-B. The result $m_{\eta} = (547.851 \pm 0.031_\mathrm{\,stat.} \pm 0.062_\mathrm{\,syst.})$ MeV agrees very well with the precise values of the NA48, KLOE and CLEO collaborations and deviates by about $5\sigma$ from the smaller, but also very precise value obtained by the GEM collaboration at COSY.
\end{abstract}

\section{Introduction}\label{sec:1}




The mass of the $\eta$ meson has been a controversial issue in recent years. Before 2000, three different experiments \cite{Dua74,Plo92,Kru95} yielded comparable masses for the $\eta$ meson. The Particle Data Group (PDG) then used these results to calculate a weighted mean mass $m_{\eta} = (547.30 \pm 0.12)$\,MeV \cite{Gro00}. In 2002, the NA48 collaboration published \cite{Lai02} a very precise result, $m_{\eta} = (547.84 \pm 0.05)$ MeV, which deviated significantly from the world average adopted by the PDG. Including the NA48 measurement in the average, the PDG in 2004 \cite{Eid04} obtained the value $m_{\eta} = (547.75 \pm 0.12)$\,MeV, almost 0.5\,MeV higher than the value reported previously. This created the motivation to repeat the previous Mainz \cite{Kru95} measurement at MAMI, especially after another precise measurement by the GEM collaboration at the COSY facility \cite{Abd05} gave the result $m_{\eta} = (547.31 \pm 0.04)$\,MeV, in agreement with the old measurements of the $\eta$ mass.

Since the previous $\eta$ mass measurement at Mainz \cite{Kru95}, the MAMI electron accelerator \cite{Her76,Wal90} was significantly improved to provide a more precisely known electron beam energy and much higher beam stability. In addition, on-line monitoring of the electron and photon beam positions in the experimental hall was introduced. Furthermore, the old TAPS setup with its limited angular coverage was replaced with the large acceptance Crystal Ball \cite{Ore82,Sta01} photon spectrometer, which allowed an improvement in the detection efficiency for the two most prominent neutral decays of the $\eta$ meson, $\eta \to 2\gamma$ (BR=39.31\%) and $\eta \to 3\pi^0$ (BR=32.56\%). Together with the high energy resolution of the tagged photon beam offered by the recently developed tagger focal-plane microscope detector \cite{Rei06}, these improvements provided the more accurate determination of the $\eta$ mass presented in this paper.

This work describes the measurement of the $\eta$ photoproduction threshold $E_{\rm thr}$ \cite{Nik11} from data measured in 2004-2005 with the Crystal Ball/TAPS detector system and the tagger focal-plane microscope detector. From kinematics of the reaction $\gamma p \to p \eta$ the $\eta$ meson mass $m_{\eta}$ was calculated using
\begin{equation}
m_{\eta}=-m_p+\sqrt{m_p^2+2\,m_p\cdot \frac{E_{\mathrm{thr}}}{c^{\,2}}},
\label{eq:emass}
\end{equation}
where $m_p$ is the proton mass and $c$ is the speed of light in vacuum.

\section{Experimental setup}\label{sec:2}

\begin{figure}
\centering
\epsfig{file=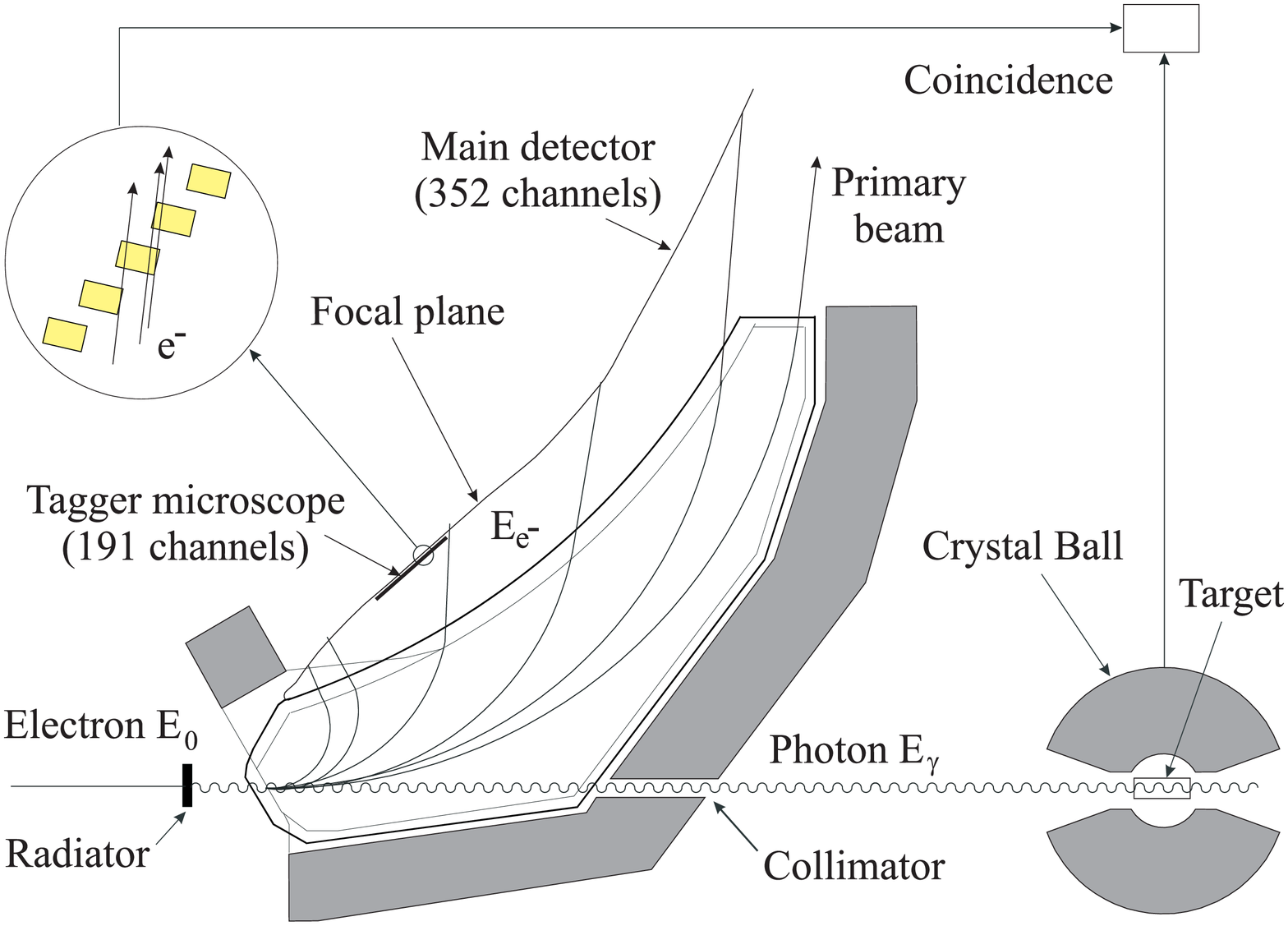,width=8.0cm}
\caption{
Plan view of the tagging bremsstrahlung facility \cite{Ant91,Hal96} and the Crystal Ball detector (not to scale) at Mainz. The tagger microscope detector \cite{Rei06}, giving improved resolution, was installed in the focal plane in front of the main detector at the position indicated. The inset shows the geometry of the microscope scintillators.}
\label{fig:tagger}
\end{figure}

In Mainz the real photon beam was produced by brems\-strahl\-ung of the 883\,MeV electrons from MAMI-B \cite{Her76,Wal90} on a 100\,$\mu$m thick diamond radiator. The absolute electron energy, $E_0$, of the incident beam was precisely determined \cite{Jan06,Jan07} in the third race-track microtron of MAMI-B with a total uncertainty of about $\sigma_{0} = 140$\,keV. The photon energies were determined using the Glasgow photon tagging spectrometer (tagger) \cite{Ant91,Hal96}, which provided a tagged photon flux of roughly $10^{5}$\,s$^{-1}\,$MeV$^{-1}$ at a beam current of about 35\,nA.


For the first time the tagger focal-plane microscope detector \cite{Rei06} was used to improve the tagged photon energy resolution. The microscope detector was placed in front of the main focal-plane spectrometer (see fig.\,\ref{fig:tagger}), so that it covered the region around the $\eta$ production threshold ($E_{\rm thr} \approx 707$\,MeV) from $E_{\gamma} = 674$\,MeV to $E_{\gamma} = 730$\,MeV at an electron beam energy $E_0 = 883$\,MeV. Made of 96 scintillator strips overlapping to one third, it provided 191 tagging channels with a higher energy resolution of about 0.29\,MeV per channel compared to approximately 1.8\,MeV available from the main focal-plane detector.

into a separate experimental area, where they can induce reactions in the experimental target. Making a coincidence between the reaction products in the experiment and the electron in the tagger, one can determine the energy of the photon $E_{\gamma}$ as the difference between the main beam energy $E_{0}$ and the electron energy $E_{\mathrm{e^-}}$ measured with the tagger $E_{\gamma}= E_0-E_{\mathrm{e^-}}$.


\begin{figure}
\centering
\epsfig{file=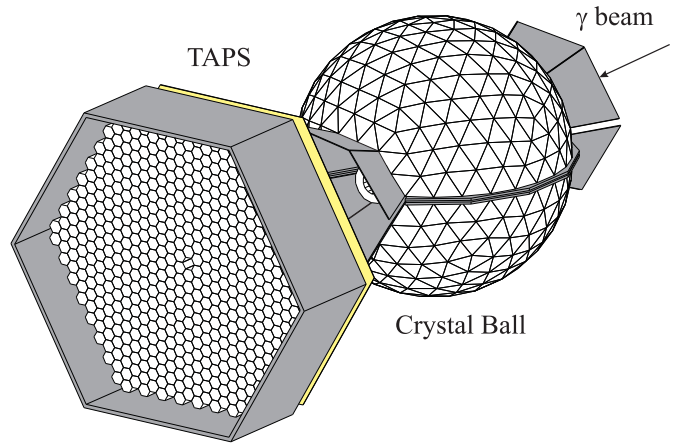,width=8.7cm}
\caption{
The Crystal Ball photon spectrometer and TAPS. The individual photomultipliers attached to the crystals are not shown; however, those supporting steel structures that were included in the Monte Carlo simulations are shown.}
\label{fig:cb}
\end{figure}

The 4.76\,cm long liquid hydrogen target was located at the center of the Crystal Ball photon spectrometer. The Crystal Ball \cite{Ore82,Sta01}, covering polar angles between $\theta = 20^{\circ}$ and $\theta = 160^{\circ}$, consisted of 672 NaI(Tl) crystals (see fig.\,\ref{fig:cb}) and had two openings for the beam in forward and backward directions. Each NaI(Tl) crystal had the form of a truncated 41\,cm long pyramid and was equipped with an individual photomultiplier. In order to distinguish between neutral and charged particles detected by the Crystal Ball, the system was equipped with a particle identification detector (PID) \cite{Wat05}. PID was a cylindrical detector, consisting of 24 2\,mm thick plastic scintillator strips, arranged parallel to the photon beam axis.

The forward wall detector, TAPS \cite{Nov91}, had 510 BaF$_2$ hexagonally shaped crystals, each equipped with a 5\,mm thick plastic scintillator for identifying charged particles. A single BaF$_2$ crystal was 25\,cm long and had an inner diameter of 5.9\,cm. The TAPS detector, intended for detecting particles in the forward direction ($\theta = 4^{\circ} - 20^{\circ}$), was located at a distance of 173\,cm from the Crystal Ball center, making it possible to use the time-of-flight analysis for the particle identification.




The experimental trigger comprised two levels. In the first level, the total energy deposited in the Crystal Ball was checked. If the sum of all photomultiplier analog signals exceeded a threshold corresponding to about 390\,MeV, the event was accepted. The second-level trigger included a condition on the Crystal Ball sector multiplicity. The 672 crystals of the spectrometer were grouped into 45 sectors of up to 16 crystals each. If at least one of the 16 signals exceeded a threshold of 20 to 40\,MeV, depending on the relative energy calibration of the photomultiplier signals, the sector contributed to the multiplicity. All events with multiplicity M\,$\geq$\,3 and every third event with M\,$\geq$\,2 were recorded for further analysis. The latter condition was especially important for detection of the $\eta \to 2 \gamma$ decay.

\section{Energy calibration of the photon beam}\label{sec:3}

\begin{figure}
\centering
\epsfig{file=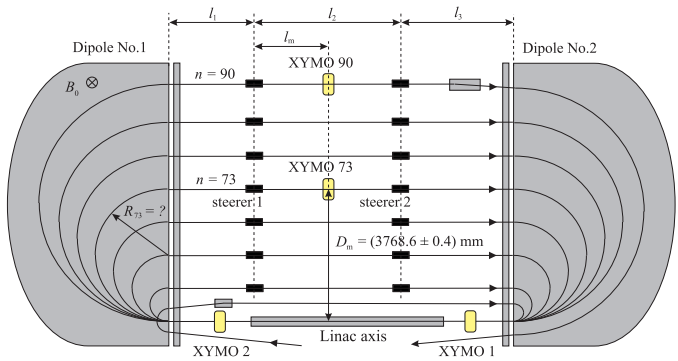,width=8.8cm}
\caption{
Determination of the bending radius $R_{73}$ of the beam in the third race-track microtron (RTM3) of MAMI-B. The position-sensitive HF monitors (XYMOs) were located \cite{Dor96} on the linac axis and on return tracks 73 and 90.}
\label{fig:rtm3}
\end{figure}

Special care was taken of the energy calibration of the tagger microscope with electrons of different known energies from MAMI. The MAMI accelerator can produce beam energies from 180\,MeV upwards in steps of 15\,MeV, and it is possible to make accurate measurements of the beam energy. Originally designed to produce electrons of maximum energy 855\,MeV, MAMI can also produce a beam of energy 883\,MeV by slightly increasing the magnetic field of the bending magnets and slightly raising the energy gain per circulation.

\begin{figure}
\centering
\epsfig{file=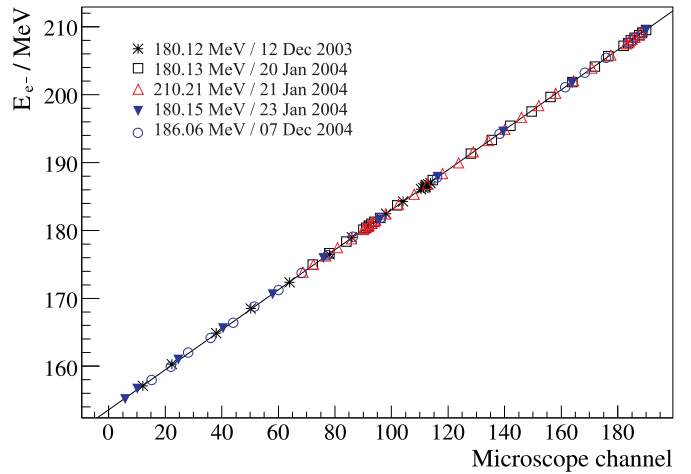,width=8.7cm}
\caption{
Tagger microscope energy calibration. Data taken from five scans with different primary MAMI beam energies are shown. Corrections for changes in the shape of the tagger magnetic field were applied. The solid line represents a linear fit to all data points.}
\label{fig:mic_cal}
\end{figure}
\begin{figure}
\centering
\epsfig{file=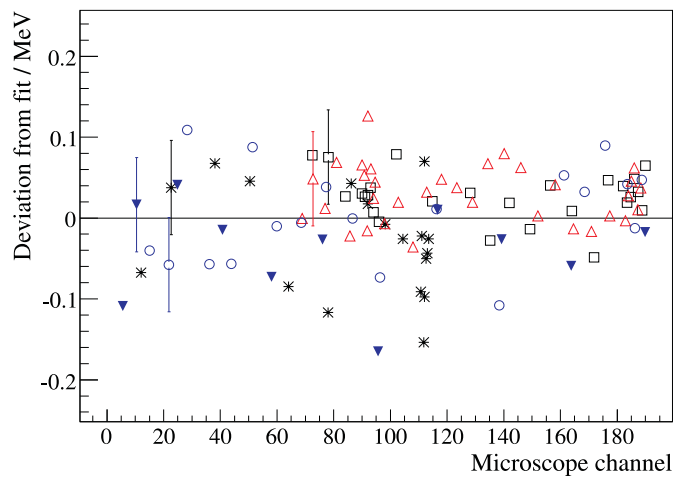,width=8.7cm}
\caption{
Tagger microscope energy calibration: deviations of the data points from the fit. The error bars represent the average RMS deviation from the fit line.}
\label{fig:mic_cal_dev}
\end{figure}
\begin{table*}
\begin{center}
\begin{tabular}{l l r} \hline
Parameter             & Procedure      & Contribution    \\\hline\hline
$E_{73}(\rm experiment)$
                      & Non-systematic MAMI uncertainty & 40\,keV \\
$E_{73}(\rm calibration)$
                      & Non-systematic MAMI uncertainty & 40\,keV \\
$\Delta E_{0}$        & PTRACE calculation              & $13 \oplus 10$\,keV \\
$\Delta E_{\rm scan}$
                      & PTRACE calculation              & $55 \oplus 20$\,keV \\
Systematic uncertainty of the calibration fit
                      & Fit parameters                  & 27\,keV \\
Initial beam misalignment
                      & MAMI optimization               & 40\,keV \\
Beam position drift in the experiment
                      & RF cavities                     & 20\,keV \\\hline
\multicolumn{2}{l}
{Total $\sigma (E_{\gamma})$}                           & 98\,keV \\\hline
\end{tabular}
\caption{
List of contributions to the total uncertainty $\sigma (E_{\gamma})$ of the photon beam energy calibration. In case the uncertainties are shown with symbol $\oplus$ (summation in quadrature), the second one originates from the fit to the PTRACE data.}
\label{tab:eg_err}
\end{center}
\end{table*}
Precise determination of the MAMI energy is im\-ple\-men\-ted in an automated manner as a standard MA\-MI operator menu \cite{Jan06,Jan07}. It is based on a precise measurement of the bending radius of the beam in dipole magnet No.\,1 (see fig.\,\ref{fig:rtm3}). After the procedure of optimization, the beam in MAMI is centered along the linac axis with the aid of X-Y position-sensitive HF monitors (XYMOs), as shown in fig.\,\ref{fig:rtm3}. By measuring the angular deviations produced by the steerer magnets and the position of the beam in track 73, with the position monitor \cite{Dor96} placed at a precisely measured distance $D_{\rm m}$ from the linac axis, the bending radius $R_{73}$ of the beam in the magnet can be found from geometrical calculations. Since the magnetic field $B_0$ of the magnets is precisely measured, the energy $E_{73}$ of the beam in returning track 73 can be calculated using
\begin{equation}
E_{73} = e\: c\: B_0\: R_{73}\,,
\end{equation}
where $e$ is the electron charge and $c$ is the speed of light in vacuum. The final energy $E_0$ of the electron beam after 90 circulations is interpolated using the calculated data from a simulation using PTRACE, a proven particle tracking program used at MA\-MI that is based on real measured magnetic field profiles.

The calibration of the tagger microscope was performed by varying the magnetic field $B_{\rm scan}$ in the tagging spectrometer around the value $B_{\rm exp}$ used in the experiment. This has been done with three different MAMI energies, $E_{\rm scan}$, to scan across the tagger microscope by increasing the value of $B_{\rm scan}$ in small steps and plotting the measured hit position of the beam in the microscope versus the equivalent energy
\begin{equation}
E_{\rm equiv} = E_{\rm scan} \cdot \frac{B_{\rm exp}}{B_{\rm scan}}.
\end{equation}
Variation of the magnetic field $B_{\rm scan}$ caused slight changes in the shape of the latter, thus, the corrections were applied for this effect using the same method as in \cite{Rei06}. The details of this method are described in \cite{Nik11}. Five calibration scans
%
%
(three made with energy $E_{\rm scan} = 180$\,MeV, one with $E_{\rm scan} = 186$\,MeV and one with $E_{\rm scan} = 210$\,MeV) gave the 107 data points that are shown in figures \ref{fig:mic_cal} and \ref{fig:mic_cal_dev}. The correlation between the data points of different scans, originating from the uncertainty of the MAMI energy, led to a block-diagonal form of the error matrix used in the fit procedure. The fit was performed by a least squares minimization with the aid of the MINUIT package, supposing a linear dependence of the tagging electron energy, $E_{e^-}$, on the microscope channel. With the known energy $E_0$ of the MAMI electron beam and the energy $E_{e^-}$ of the tagging electron, the energy $E_{\gamma}$ of the bremsstrahlung photon is determined using
\begin{equation}
E_{\gamma} = E_0 - E_{e^-}.
\label{eq:egamma}
\end{equation}
The determination of $E_{0}$ was made four times during the experiment with the average value
\begin{equation}
E_{0} = (883.057 \pm 0.134_{\rm\,syst.} \pm 0.040_{\rm\,non\textrm{-}syst.})\,\mathrm{MeV}.
\end{equation}

For tagging electron energies equal to the values of $E_{\rm scan}$, the photon energy $E_{\gamma}$, given by eq.\,(\ref{eq:egamma}), is calculated as the difference between two MAMI energies. Thus, any systematic uncertainty in determination of $E_{73}$ cancels and the non-systematic part contributes twice:
\begin{equation}
E_{\gamma} = E_{73}(\mathrm{exp.}) + \Delta E_{0} - E_{73}(\mathrm{cal.}) - \Delta E_{\rm scan}\,,
\end{equation}
where $E_{73}$ (exp.) is the energy of the experiment, $E_{73}$ (cal.) was measured once for each of the calibration scans, and $\Delta E_{0}$ and $\Delta E_{\rm scan}$ are the differences (calculated by PTRACE) between the measured energy $E_{73}$ and the output MAMI energies $E_0$ and $E_{\rm scan}$, respectively. The main contributions to the total uncertainty of the MAMI energy, $\sigma_0 = 140$\,keV, are the uncertainty of the distance $D_{\rm m}$ (see fig.\,\ref{fig:rtm3}), measured by the geodetic method, which is a systematic contribution, and the measured magnetic field uniformity of the MAMI magnets. Since the PTRACE simulation already includes the measured profile of the magnetic field, this contribution is also systematic, and the uncertainty due to the measurement of the absolute magnetic field $B_0$, made with an NMR system, can be neglected. The non-systematic contribution, due to the uncertainty in the beam position and in the measurement of the angular deviations produced by the steerer 
magnets, was estimated to be about 38\,keV. In order to obtain an objective estimate of the non-systematic contribution, the RMS deviation from the average of the 106 values of the MAMI energy $E_{73}$ measured in the period from 04/2004 to 07/2009, was calculated. The calculation resulted in $\sigma_0$(non-syst.) $=40$\,keV.

All contributions to the uncertainty of the photon beam energy calibration are summarized in table \ref{tab:eg_err}. The values of $\Delta E_0$ and $\Delta E_{\rm scan}$ were calculated by PTRACE with uncertainties 13\,keV and 55\,keV \cite{Jan06}, respectively. The systematic uncertainty of the calibration fit of 27\,keV ($\sigma$) was obtained from the fit parameters by applying the law of error propagation to the linear fit function.
%
%
The initial misalignment of the electron beam at radiator at the beginning of the scans introduced the uncertainty of about $40$\,keV ($\sigma$) \cite{Nik11}, and the contribution caused by the drift of the beam position during the experiment was estimated to be $20$\,keV ($\sigma$) \cite{Nik11}. Added in quadrature, all contributions resulted in a total uncertainty of $\sigma(E_{\gamma}) = 98$\,keV in the determination of the photon energy $E_{\gamma}$. Applying the law of error propagation to eq.\,(\ref{eq:emass}), the systematic uncertainty of the $\eta$ mass due to photon beam energy calibration was estimated to be $\sigma(m_{\eta}) = 62$\,keV.

\section{Analysis and results}\label{sec:4}

The $\eta$ mesons were identified via their two main decay modes, $\eta \to 2\gamma$ and $\eta \to 3\pi^0$, with the Crystal Ball/TAPS setup, which measured energies and emission angles of particles. The energy calibration of the Crystal Ball was performed by identifying the reaction $\gamma p \to p \pi^0$. The gains of the photomultipliers were adjusted for all crystals so that the peak position in the two photon invariant mass distribution agreed with the well-known $\pi^0$ mass. A cluster in the Crystal Ball was formed by a group of adjacent crystals that had registered parts of the electromagnetic shower initiated by a particle. The weighted mean of the vectors of all contributing crystals, using the square root of the energy in each cluster element as weight, was taken as the vector of the cluster, and the energy sum of these elements gave the total cluster energy. Only crystals with energy deposits greater than 2\,MeV could contribute to a cluster, and the total energy threshold for clusters was set to 
20\,MeV. The charged clusters were identified with PID by checking the agreement of the azimuthal angles of Crystal Ball clusters with the angles of hit PID elements. The uncharged clusters were considered to be photons. In the first step of the analysis, events with two and six coincident photons were selected to pick out candidate $\eta$ events from the two decay modes mentioned above.

\begin{figure}
\centering
\epsfig{file=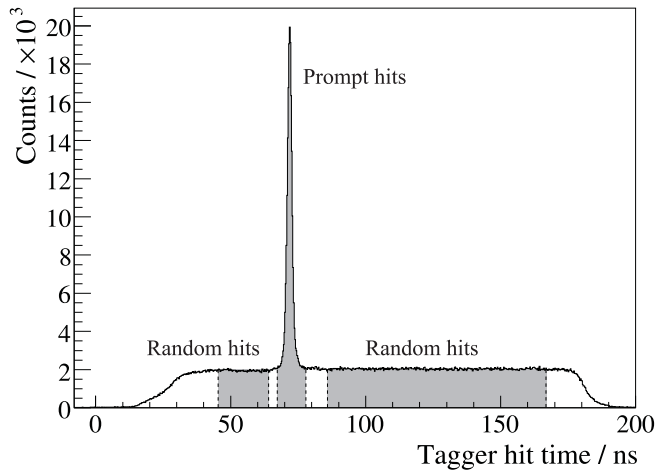,width=8.5cm}
\caption{
Timing distribution of the tagger hits. The $x$-axis shows the relative time of the tagger hit with respect to the average time of the Crystal Ball event.}
\label{fig:tag_time}
\end{figure}

The high intensity of the electron beam caused, for each event, several electron hits to be registered in the tagger focal-plane detector or in the tagger microscope. Only one of these electrons could be correlated to the photon that induced the reaction. This ambiguity was resolved using a coincidence analysis. Since the involved electron had a fixed time difference to the event trigger produced by the Crystal Ball, the tagger hits were cut using a time window around the coincidence peak in the tagger time distribution (prompt time window) shown in fig.\,\ref{fig:tag_time}. The contribution caused by random coincidences was subtracted using the tagger hits within an additional time window in the region of random hits (random time window).

%

\begin{figure}
\centering
\epsfig{file=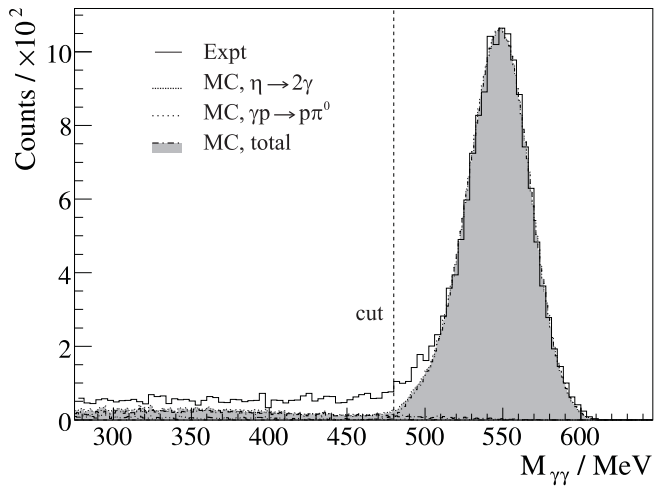,width=8.5cm}\vspace{0.20cm}
\epsfig{file=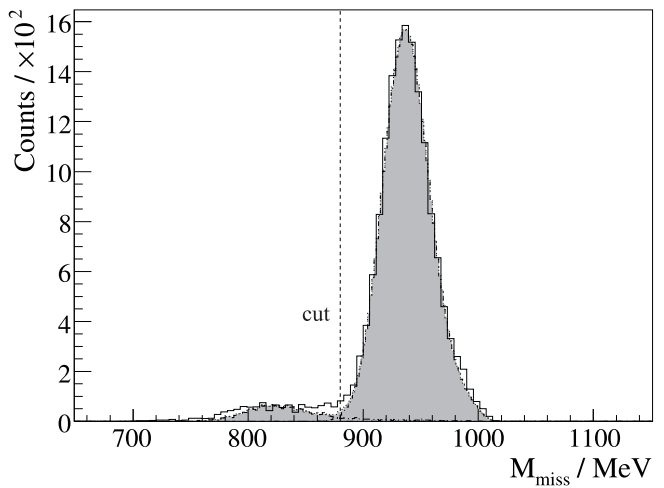,width=8.5cm}
\caption{
Top: two-photon invariant mass distribution for $707 < E_{\gamma} < 730\,\mathrm{MeV}$ after a cut on the missing mass. Bottom: two-photon missing mass distribution after a cut on the invariant mass. The dashed histograms were produced by Monte Carlo simulation.}
\label{fig:eta_invm2g}
\end{figure}
\begin{figure}
\centering
\epsfig{file=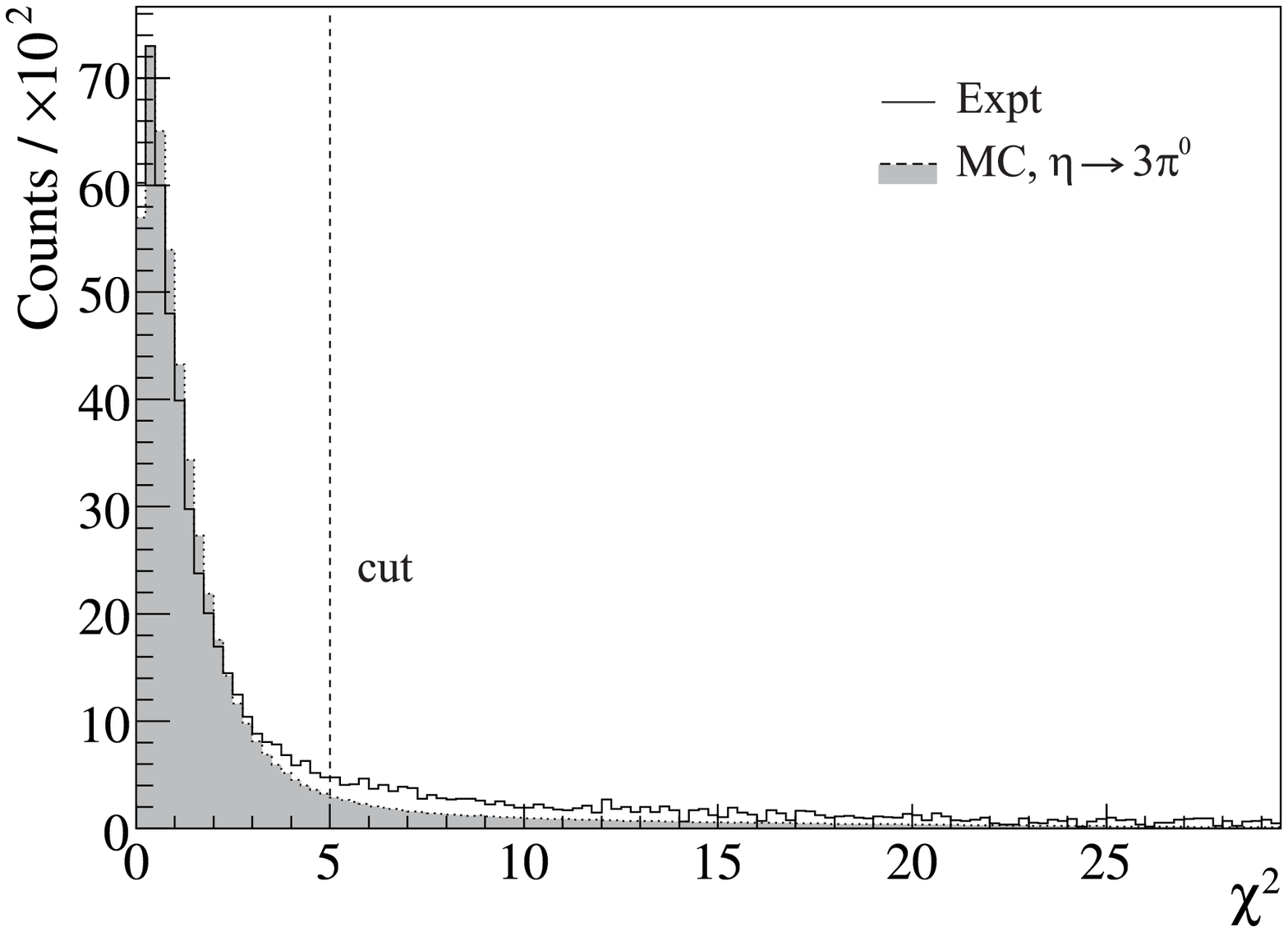,width=8.5cm}\vspace{0.35cm}
\epsfig{file=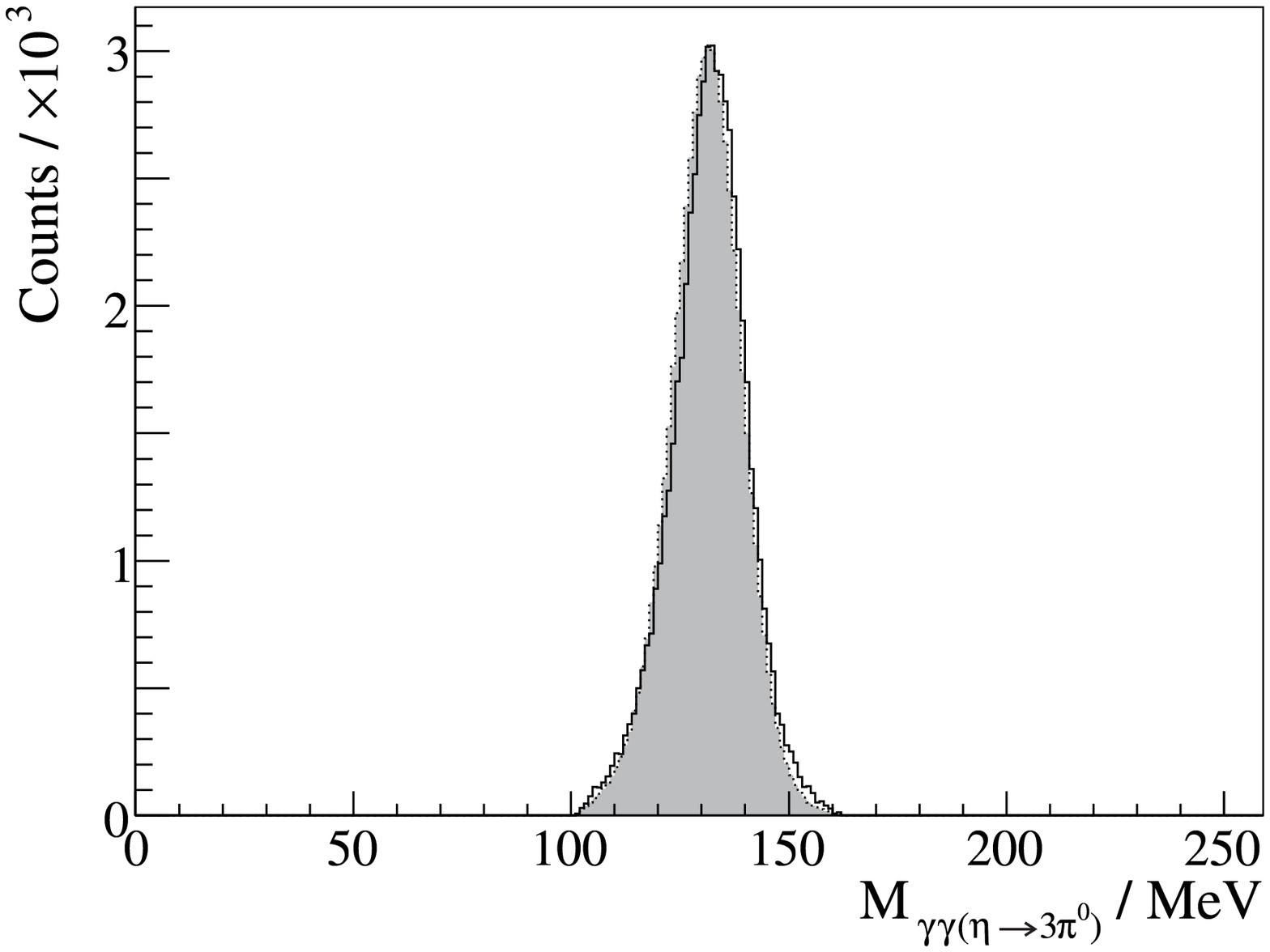,width=8.5cm}
\caption{
Top: distribution of the minimum $\chi^2$ from $6\gamma$-events for $707 < E_{\gamma} < 730\,\mathrm{MeV}$. Bottom: invariant mass distribution of photon pairs from $6\gamma$-events after applying a cut at $\chi^2 < 5$. The dashed histograms indicate Monte Carlo simulation of the $\eta \to 3\pi^0$ events.}
\label{fig:eta_invmP0}
\end{figure}
The identification of the $\eta \to 2\gamma$ decay concentrated on events with a trigger multiplicity M\,$\geq$\,2 and two clusters detected as photons, ignoring all other particles. The standard invariant mass analysis using
\begin{equation}
M_{\gamma \gamma} = \sqrt{(E_1+E_2)^2-(\vec p_1+\vec p_2)^2}
\end{equation}
with 4-vectors $(E_{1,2},\,\vec p_{\,1,2})$ of the two identified photons showed a peak at the $\eta$ mass with a resolution of $\sigma \approx 20$\,MeV. Using the tagger hits in the prompt and random time windows, the distribution of the missing mass $M_{\mathrm{miss}}$ of the undetected particle (proton) was produced (see fig.\,\ref{fig:eta_invm2g}). Cuts were applied on the invariant mass at $M_{\gamma\gamma}>480$\,MeV and on the missing mass at $M_{\mathrm{miss}}>880$\,MeV. The two-photon invariant mass $M_{\gamma \gamma}$ distribution after the cut on the missing mass and subtraction of the random tagger hits is shown in fig.\,\ref{fig:eta_invm2g} as the solid histogram. The combinatorial background at smaller invariant masses arises mainly from $\pi^0$ production, for example, if two photons from different $\pi^0$ mesons are detected within the time resolution or if $\pi^0\pi^0$ events are produced with two escaping photons. The invariant and missing mass distributions of the $\eta \to 2\gamma$ decay 
were simulated using a code based on the GEANT 3.21 simulation library, including all features of the target and detector setup. The simulation of $\pi^0$ and $\pi^0\pi^0$ photoproduction showed that the contribution of the background to the $\eta$ meson candidates was less than 1.5\%. Simulated $\eta$ events generated an invariant mass in good agreement with the measured data. Though the agreement between simulated and measured background was not perfect, the inconsistency did not extend above $M_{\gamma\gamma} \approx 480$\,MeV. After subtraction of the background caused by the random coincidences between the Crystal Ball and the tagger microscope, the $\eta$ yield below production threshold was very close to zero.

\begin{figure}
\centering
\epsfig{file=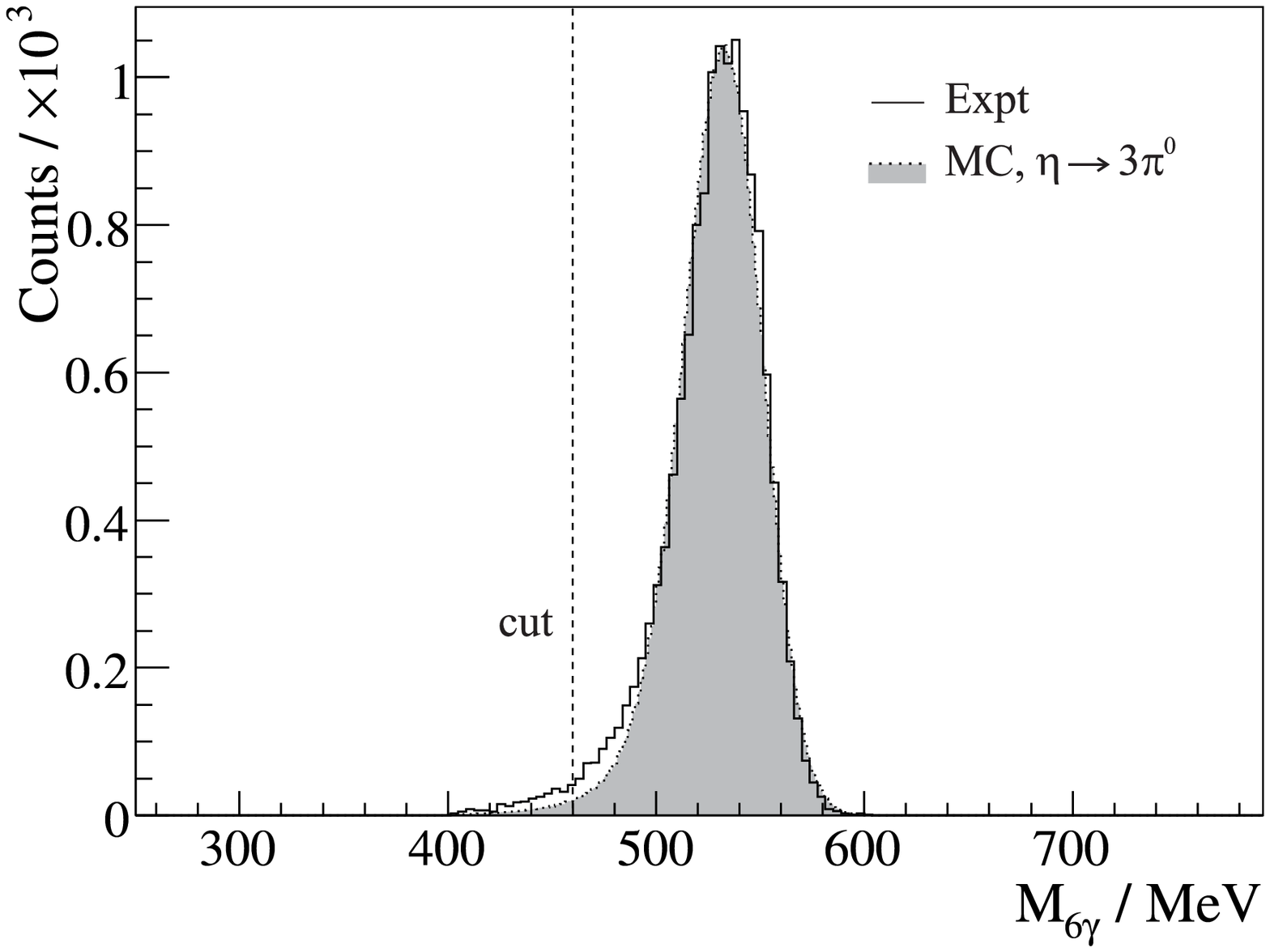,width=8.5cm}\vspace{0.15cm}
\epsfig{file=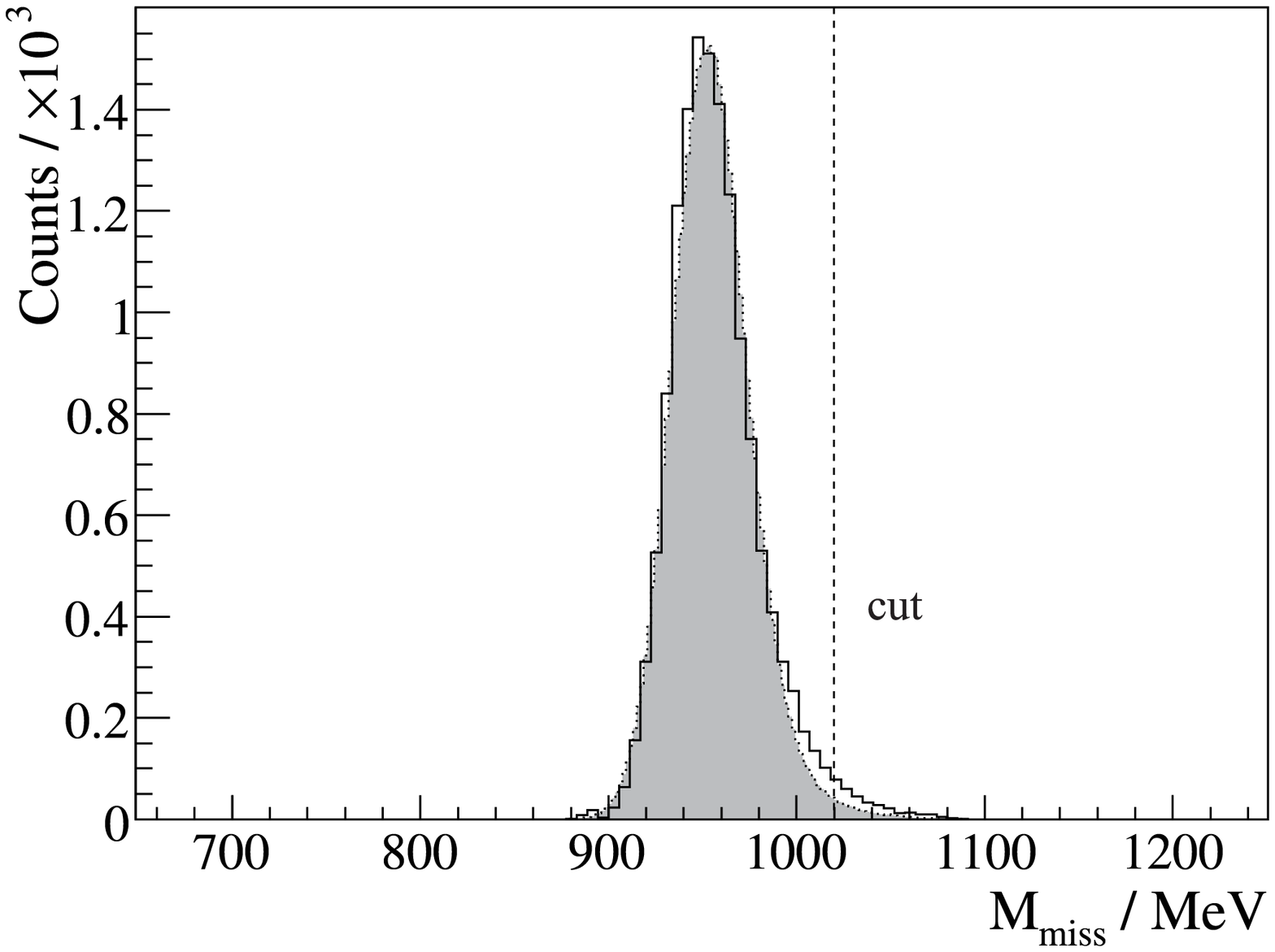,width=8.5cm}
\caption{
Top: six-photon invariant mass distribution for $707 < E_{\gamma} < 730\,\mathrm{MeV}$ after applying the $\chi^2 < 5$ cut of fig.\,\ref{fig:eta_invmP0}. Bottom: six-photon missing mass distribution. The dashed histograms indicate Monte Carlo simulation of the $\eta \to 3\pi^0$ events.}
\label{fig:eta_invm6g}
\end{figure}

The identification of the $\eta \to 3\pi^0 \to 6\gamma$ decay concentrated on events with a trigger multiplicity M\,$\geq$\,3 and six clusters detected as photons, ignoring all other particles. Among the 15 possible combinations of six photons to be arranged in three pairs, the combination with the minimum $\chi^2$-value:
%
%
%
\begin{equation}
\chi^2 = \frac{1}{3 \sigma_{\gamma\gamma}^2} \sum_{i=1}^{3}{(m_{\gamma\gamma}[i]-m_{\pi^0})^2}
\label{eq:chi2}
\end{equation}
was assumed to be correct. Here $m_{\gamma\gamma}[i]$ are the invariant masses of the photon pairs, $\sigma_{\gamma\gamma}$ is the width of the invariant mass distributions and $m_{\pi^0}$ is the well-known $\pi^{0}$ mass. Figure \ref{fig:eta_invmP0} shows the distribution of the minimum $\chi^2$ and distribution of the photon pairs invariant mass after a cut at $\chi^2$\,$<$\,$5$ was applied. Simulation of the channel with the GEANT code showed that 12\% of the simulated $\eta \to 3\pi^0$ events generated a $\chi^2$ beyond this threshold. The invariant mass and missing mass distributions of the six photons, shown in fig.\,\ref{fig:eta_invm6g}, were in quite good agreement with the simulation. Additional cuts on the invariant mass at $M_{6\gamma} > 460$\,MeV and on the missing mass $M_{\rm miss} < 1020$\,MeV were applied.

The main background contribution was caused by the direct $3\pi^0$ production through the reaction $\gamma p \to 3\pi^0 p$. Since below the $\eta$ production threshold ($E_{\gamma} < 707$\,MeV) no other process can produce six or more photons, the contribution of the resonant $3\pi^0$ production can be estimated by measuring the $\eta$ cross section below threshold, which ideally must be zero. Such an estimate resulted in approximately 0.12 a.\,u. (see fig.\,\ref{fig:eta_total}) Supposing that the considered process made the same contribution to the $\eta$ cross section above the $\eta$ threshold, the contribution of the $3\pi^0$ events to the $\eta$ events can be estimated to be about 2\% in the energy region $707 < E_{\gamma} < 730\,\mathrm{MeV}$.


%
%

\subsection{Total cross section}

\begin{figure}
\centering
\epsfig{file=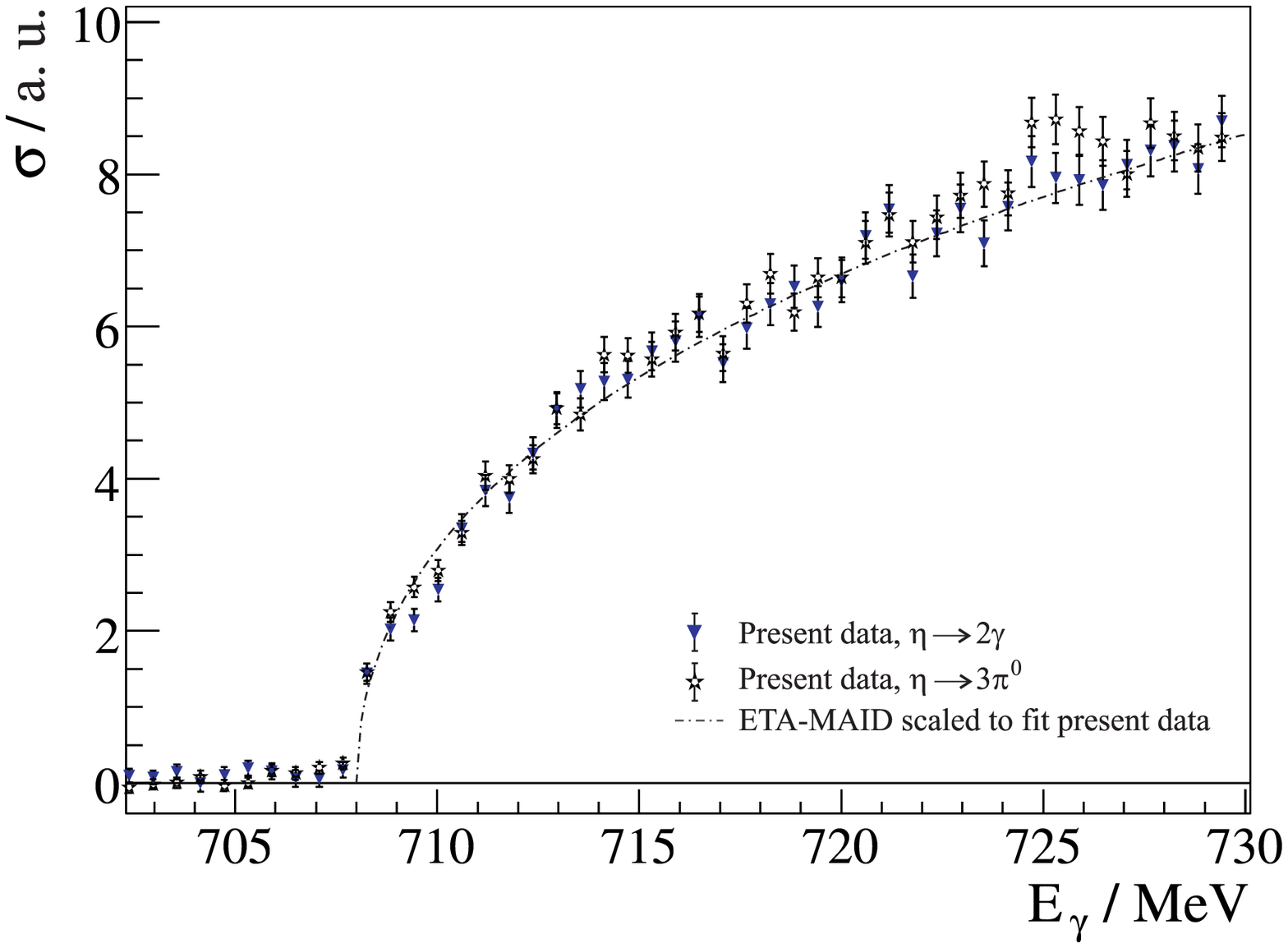,width=8.4cm}\vspace{0.2cm}
\epsfig{file=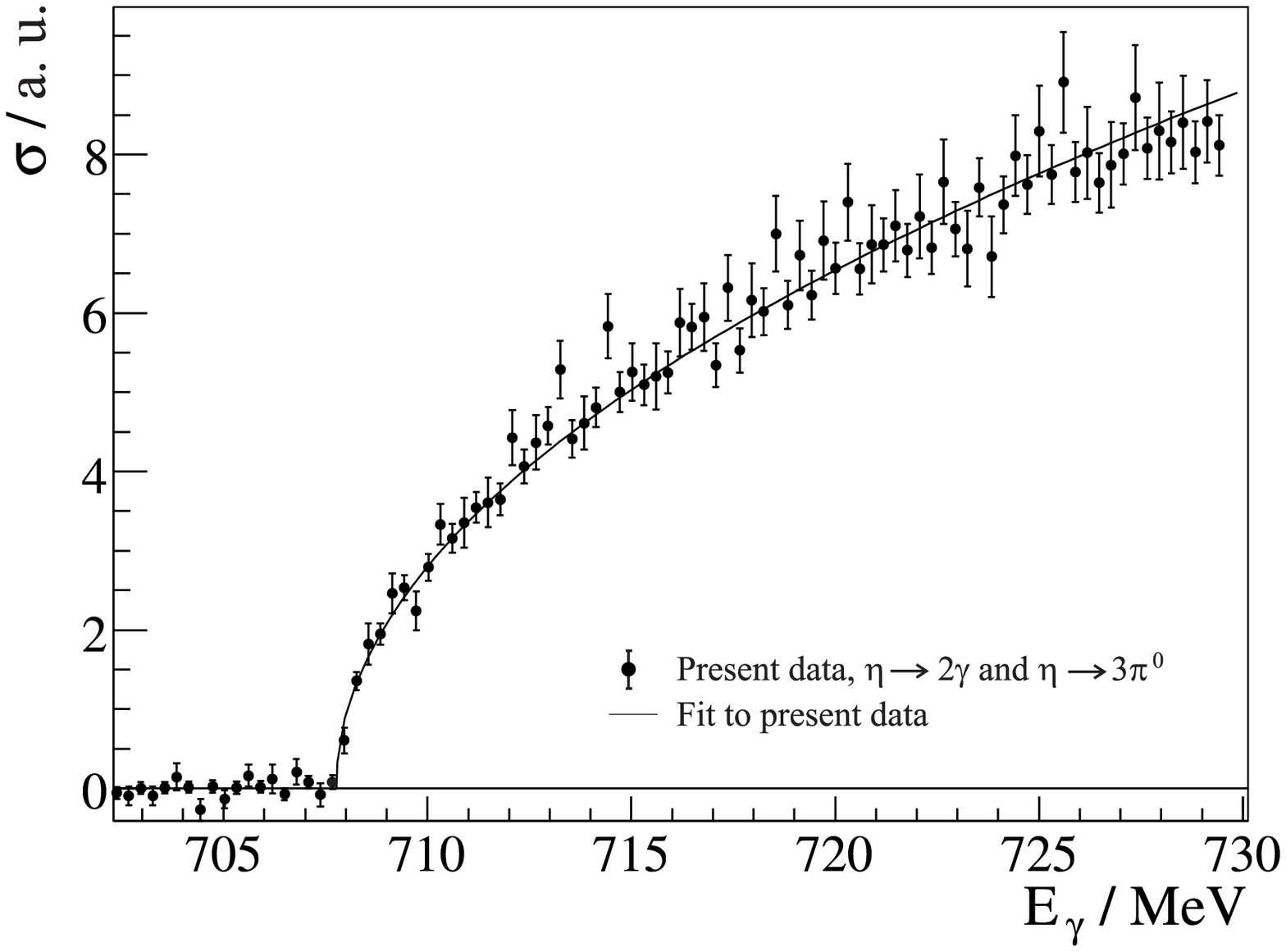,width=8.4cm}
\caption{
Total cross section for the $\gamma p \to p \eta$ reaction in arbitrary units. Top: cross sections obtained by separate analysis of data from the two $\eta$ decay modes shown in 0.59\,MeV steps (microscope strips) in comparison to the scaled prediction of the ETA-MAID isobar model \cite{Chi02}. Bottom: cross section obtained by summing all $\eta$ events shown in 0.29\,MeV steps (microscope channels).}
\label{fig:eta_total}
\end{figure}



The normalization for the total cross section was obtained from the intensity of the photon flux, the simulated acceptance of the Crystal Ball/TAPS, and branching ratios of the $\eta$ decays. The photon flux was determined by counting electrons detected in the tagger, and measuring the ratio of the number of tagged photons passing to the experimental area, to the number of the tagger electrons (tagging efficiency). The acceptance of the Crystal Ball was determined by analysis of the events simulated with the GEANT code. For the analysis described in this paper, $10^7$ events for each of the two considered $\eta$ decay modes were generated in the range $707 < E_{\gamma} < 730$\,MeV. The acceptance of about 25.2\% at the $\eta$ production threshold, smoothly decreasing to about 24.6\% at $E_{\gamma} = 730$\,MeV, was obtained for the decay $\eta \to 2 \gamma$. The analysis of the $\eta \to 3 \pi^0$ decay gave the acceptance of approximately 38.5\% at the $\eta$ threshold, smoothly decreasing to about 36.8\% at 
$E_{\gamma} = 730$\,MeV.

The resulting total $\eta$ cross section is shown in fig.\,\ref{fig:eta_total} in arbitrary units (a.\,u.). As a first step, the total cross section was obtained using the microscope strips and is shown in fig.\,\ref{fig:eta_total} top in 0.59\,MeV steps for the two considered decays of the $\eta$ meson. The absolute cross sections obtained using the $\eta \to 2\gamma$ and $\eta \to 3 \pi^0$ decays agree well with each other, and the shape of the cross section is in good agreement with the prediction of the ETA-MAID isobar model \cite{Chi02}. Though the background caused by the target windows was measured with an empty target and subtracted, it was not possible to avoid some negligible residual background ($\approx$\,0.12 a.\,u. in average) below threshold, caused by the background processes considered in section \ref{sec:4}. The plot in fig.\,\ref{fig:eta_total} bottom shows the cross section obtained by summing all $\eta$ events with the full resolution of the microscope of 0.29\,MeV per channel. In order 
to simplify the procedure of the threshold energy determination, the background below threshold was linearly fitted and extrapolated into the $\eta$ region. In fig.\,\ref{fig:eta_total} bottom, this background has been subtracted which decreased the cross section by about 2\%. The solid line represents the fit to the data with a function considered in the next section.





\subsection{Threshold energy and the \boldmath $\eta$ \unboldmath mass}

Since the determination of the $\eta$ mass requires a very precise measurement of the production threshold, it was necessary to determine the behavior of the cross section near threshold. Due to strong dominance of the S$_{11}$(1535) resonance in the threshold region \cite{Kru95a} it is expected for the total cross section
\begin{equation}
\sigma(E_{\gamma}) \propto (E_{\gamma}-E_{\mathrm{thr}})^{1/2}.
\end{equation}
%
%
%
The function based on such dependence,
\begin{equation}
f(E_{\gamma}) = a_1\,(E_{\gamma}-E_{\mathrm{thr}})^{1/2},
\label{eq:func_sigm}
\end{equation}
gave good agreement with the shape of the cross section and was fitted to the total $\eta$ cross section to determine the threshold energy $E_{\rm thr}$. To improve the statistical uncertainty it was decided to fit the cross section obtained by summing the reconstructed events of both decays.

\begin{figure}
\epsfig{file=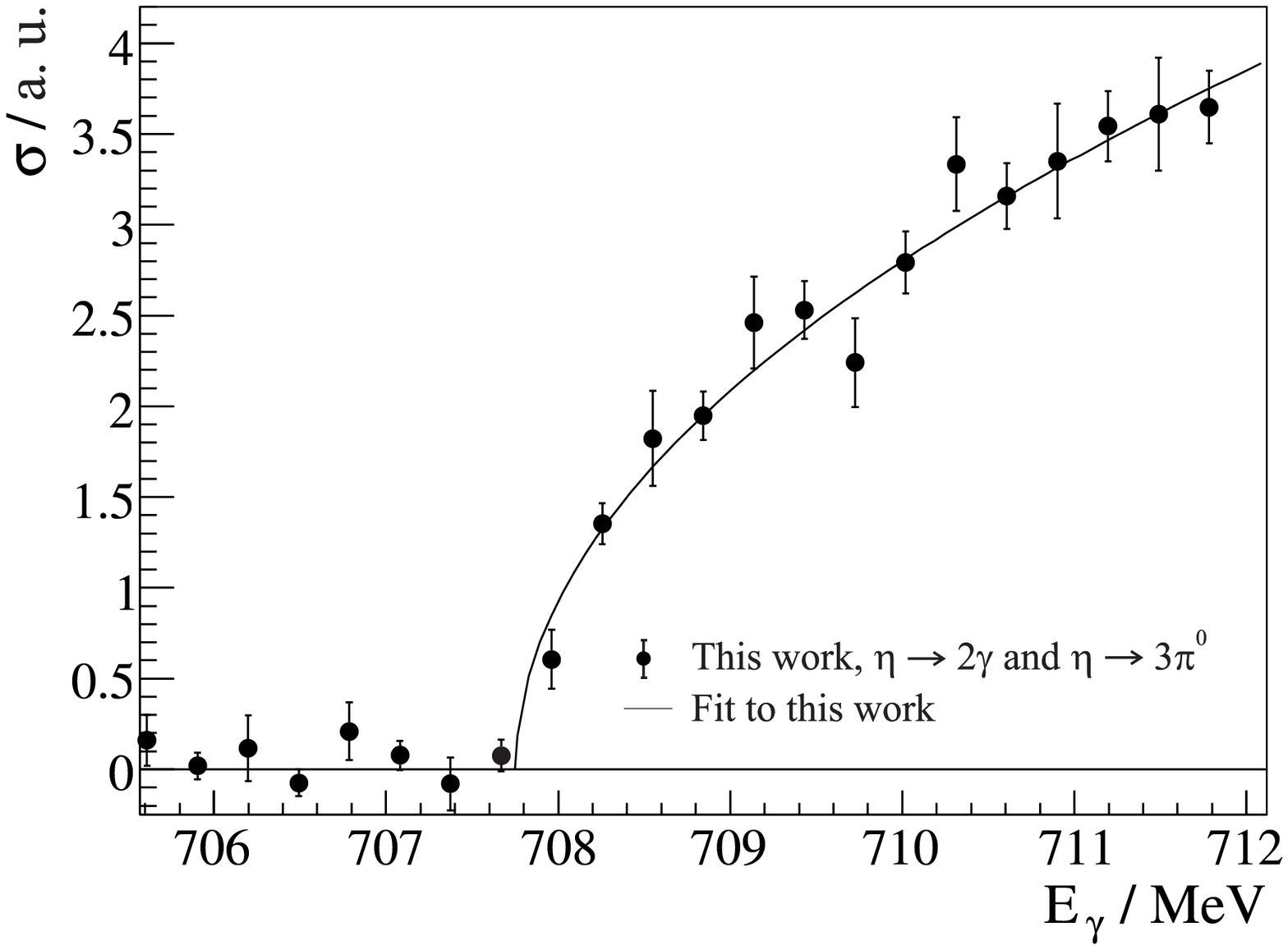,width=7.70cm}\vspace{0.2cm}
\epsfig{file=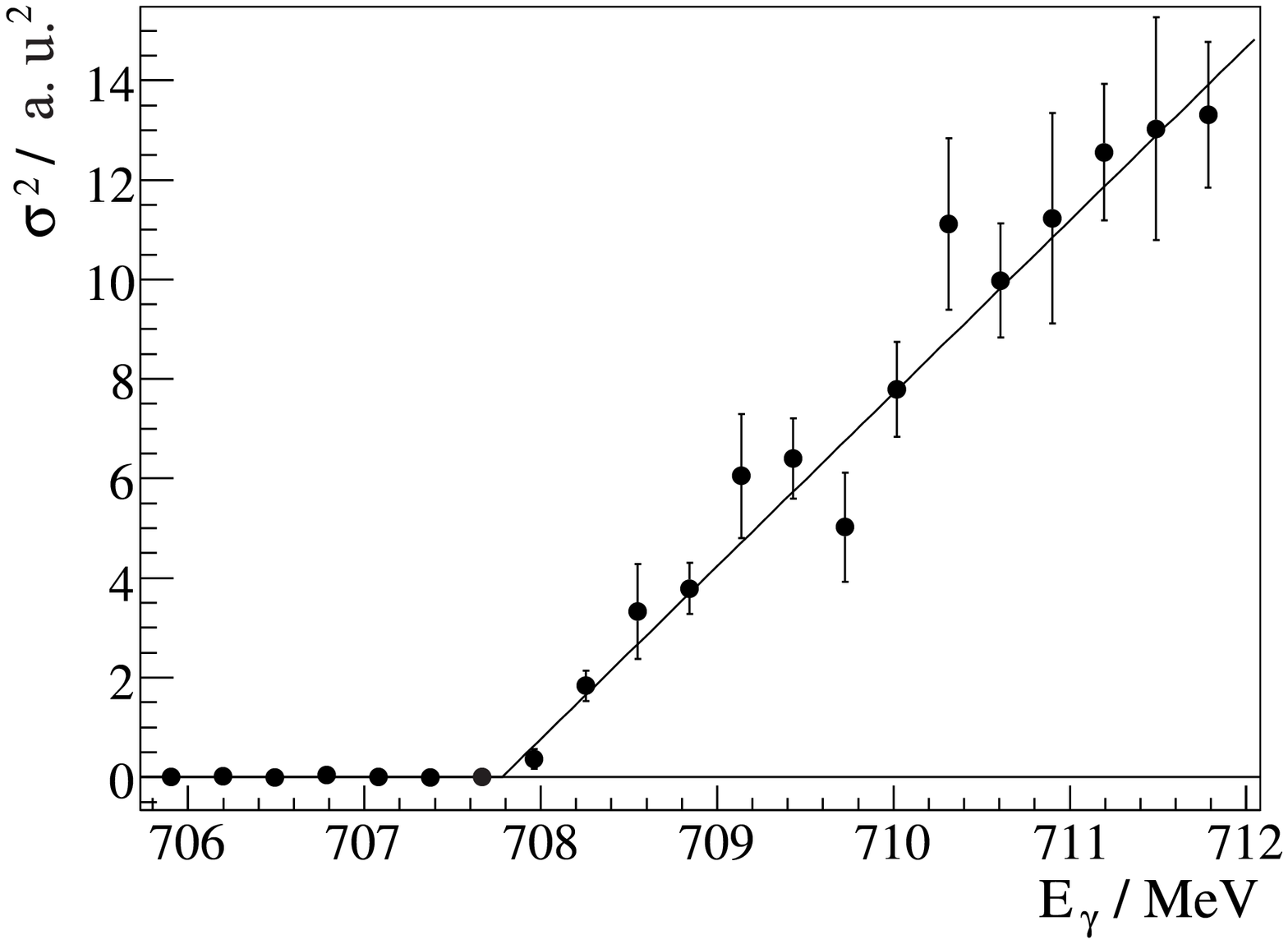,width=7.75cm}\vspace{0.2cm}
\epsfig{file=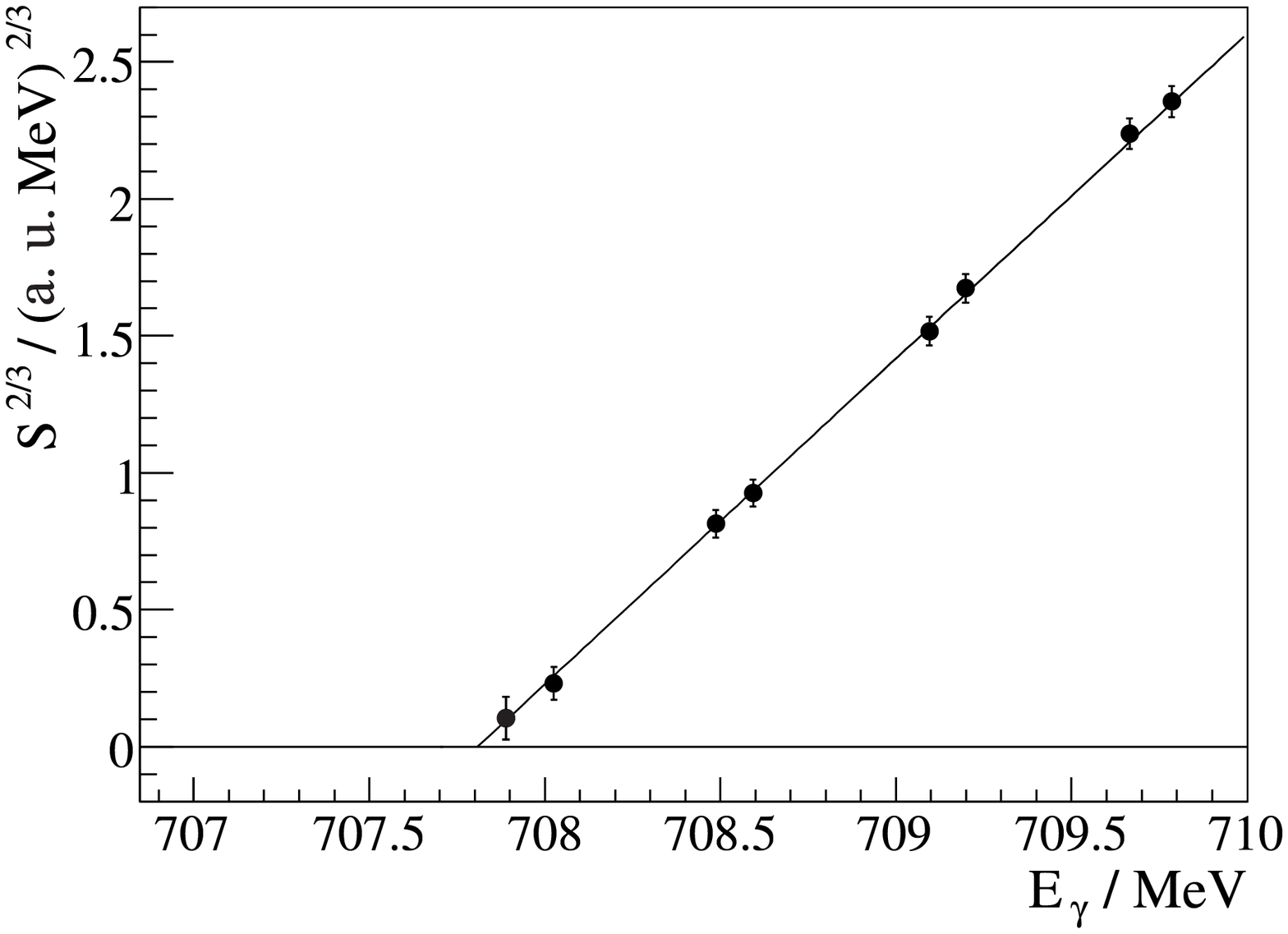,width=7.83cm}
\caption{
The total $\eta$ cross section $\sigma$, the square of the cross section $\sigma^2$ and the integrated cross section $S^{2/3}$ near production threshold. Solid lines represent fits for determination of the threshold energy.}
\label{fig:thres}
\end{figure}

Due to the finite size of the energy bins, the number of counts in the center of the bin is proportional to the average value of the cross section over the width of the bin, so that its content will not be zero even if the bin only slightly overlaps the threshold. Therefore, the center of the first bin with non-zero counts can be below the threshold, which makes the fitting procedure unstable. One possibility to take this effect into account and make the fitting procedure more reliable is not to use this bin in the fit. Another possibility, which was suggested in \cite{Kru95}, is to fit the integral of the cross section. The experimental value of the integrated cross section is given by
\begin{equation}
S(E_{\gamma}) = \sum_{i=n_0}^{n} \sigma(i)\cdot \Delta(i),
\end{equation}
where $\sigma(i)$ is the cross section in the energy bin $i$, and $\Delta(i)$ is the width of the bin (the odd and even bins have different width). The sum starts at $n_0$, which is the first bin with non-zero counts, and gives the value of $S$ at the upper edge of the bin $n$, which corresponds to the photon energy
\begin{equation}
E_{\gamma} = E_{\gamma}(n)+\frac{\Delta(n)}{2}.
\end{equation}
Due to $s$-wave-like energy behavior of the cross section it is expected that
\begin{equation}
S(E_{\gamma}) \propto (E_{\gamma}-E_{\rm thr})^{3/2},
\end{equation}
and therefore $S^{2/3}$ should be a linear function of the photon energy $E_{\gamma}$.



\begin{figure}
\centering
\epsfig{file=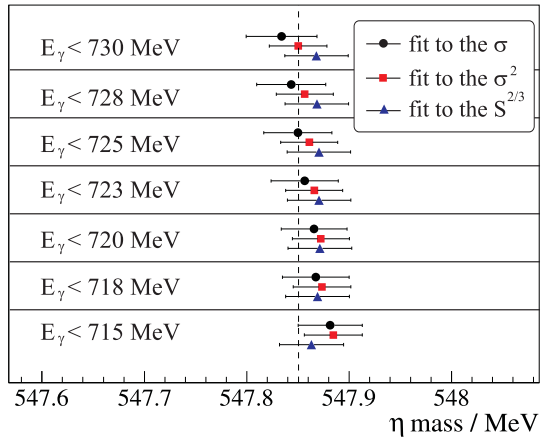,width=7.0cm}
\caption{
The $\eta$ mass obtained using different methods to fit the threshold energy plotted versus the fit range. The dashed line indicates the result of this work.}
\label{fig:range}
\end{figure}

The threshold energy was obtained by fitting the cross section $\sigma$, the square of the cross section $\sigma^2$, and the integral $S^{2/3}$. The $\sigma^2$ was fitted with a linear function. The integral $S^{2/3}$ showed almost linear behavior close to threshold, but at higher energies a small quadratic term appeared \cite{Kru95}; therefore, the $S^{2/3}$ data were fitted with a second order polynomial. The fits are shown in fig.\,\ref{fig:thres}. They delivered consistent values for the threshold energy, which were converted to $\eta$ mass using eq.\,(\ref{eq:emass}) and are summarized in table \ref{tab:res}. The values, derived separately from the $\eta \to 2\gamma$ and $\eta \to 3\pi^0$ events, also agree well within the statistical uncertainties. Analysis of the additional experimental data of the MDM experiment \cite{Sch10} also led to a consistent result for the $\eta$ mass. Figure \ref{fig:range} shows the $\eta$ mass obtained by fitting $\sigma$, $\sigma^2$, and $S^{2/3}$, plotted versus the 
upper limit of the fit range, showing good agreement with the $\eta$ mass found by fitting over the full $E_{\gamma}$ range up to 730\,MeV.

\begin{table*}
\begin{center}
\begin{tabular}{l c c c} \hline
& $m_{\eta}$ [MeV], fit $\sigma$
& $m_{\eta}$ [MeV], fit $\sigma^2$
& $m_{\eta}$ [MeV], fit $S^{2/3}$  \\\hline\hline
%
%
$\eta \to 2\gamma+\eta \to 3\pi^0$ & $547.834 \pm 0.035$ & $547.851 \pm 0.028$ & $547.869\pm 0.031$ \\
$\eta \to 2\gamma$                 & $547.826 \pm 0.051$ & $547.837 \pm 0.046$ & $547.824\pm 0.048$ \\
$\eta \to 3\pi^0$                  & $547.822 \pm 0.044$ & $547.853 \pm 0.033$ & $547.860\pm 0.038$ \\
$\eta \to 3\pi^0$ (MDM exp.)       & $547.843 \pm 0.058$ & $547.849 \pm 0.055$ & $547.861\pm 0.056$ \\\hline
\end{tabular}
\caption{
Comparison of the results for the $\eta$ mass obtained by fitting the total cross section $\sigma$ and $\sigma^2$, and the integral $S^{2/3}$ up to $E_{\gamma} = 730$\,MeV. The most precise result was obtained by fitting all of the data, but results from fitting the $\eta \to 2\gamma$ and $\eta \to 3\pi^0$ data separately are also listed. The uncertainties are statistical only.}
\label{tab:res}
\end{center}
\end{table*}

\begin{figure}
\centering
\epsfig{file=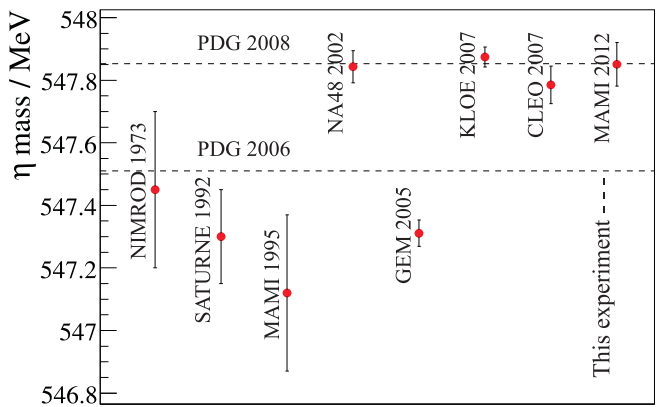,width=8.5cm}
\caption{
Overview of previous $\eta$ mass measurements in comparison to the world average reported by the Particle Data Group \cite{Yao06,Ams08} in 2006 and 2008 and the result of this analysis\,\cite{Nik11}.}
\label{fig:prev_eta_mass}
\end{figure}

The average of the three values listed in the first row of table \ref{tab:res} resulted in the $\eta$ mass
\begin{equation}
m_{\eta}=(547.851 \pm 0.031_{\mathrm{\,stat.}} \pm 0.062_{\mathrm{\,syst.}})\,\textrm{MeV,}
\end{equation}
where the first uncertainty is due only to statistics, and the second originates from the uncertainty in the photon beam energy calibration. The result of this work supports the three most precise measurements by the NA48 \cite{Lai02}, KLOE \cite{Amb07} and CLEO \cite{Mil07} collaborations and disagrees by about $5\sigma$ with the smaller value obtained by the GEM \cite{Abd05} collaboration. The $\eta$ mass determined in this work is plotted in fig.\,\ref{fig:prev_eta_mass} with the other measurements in the order of the year of publication. The disagreement with the previous measurement \cite{Kru95} at MAMI is most probably due to
%
%
the lack of online beam monitors that resulted in underestimation of the systematic uncertainty for the photon tagger energy calibration.

\section{Conclusion}\label{sec:5}

This article describes the determination of the $\eta$ mass by measuring the threshold of the reaction $\gamma p \to p \eta$ at the MAMI accelerator. The high resolution tagger microscope was used for the first time to get the total cross section of the reaction. The $\eta$ mesons were selected by identifying the decay products of the two most prominent neutral decays, $\eta \to 2\gamma$ and $\eta \to 3\pi^0$, in the Crystal Ball detector. The three precise $\eta$ mass measurements by the NA48, KLOE and CLEO collaborations were confirmed, though the result presented in this article disagrees with the GEM collaboration measurement. The uncertainty for the new $\eta$ mass measurement has been improved in comparison to the previous Mainz experiment by a factor of $\sim$\,$3$. The disagreement with the previous MAMI measurement is most probably due to the underestimated systematic uncertainty of the old result.
\\
\\
\textit{Acknowledgments:} The authors wish to thank the accelerator group of MAMI for the precise and very stable beam conditions. This work was supported by the Deutsche For\-schungs\-ge\-mein\-schaft (SFB 443, SFB/TR 16), the DFG\--RFBR (Grant No. 05-02-04014), European Community-Research Infrastructure Activity under the FP6 ``Structuring the European Research Area Programme" (HadronPhysics, Contract No. RII3-CT-2004-506078), the NSERC (Canada), Schweizerischer Nationalfonds, U.K. EPSRC, PrimeNet network, and the U.S. DOE and U.S. NSF. We thank the undergraduate students of Mount Allison University and The George Washington University for their assistance.


\end{document}